\documentclass[conference]{IEEEtran}
\usepackage{amsmath,amssymb,amsfonts}
\usepackage{algorithmic}
\usepackage{graphicx}
\usepackage{textcomp}
\usepackage{xcolor}
\usepackage{url}
\usepackage{float}
\def\BibTeX{{\rm B\kern-.05em{\sc i\kern-.025em b}\kern-.08em
    T\kern-.1667em\lower.7ex\hbox{E}\kern-.125emX}}

\usepackage[]{natbib}
\usepackage{booktabs}
\usepackage{mdwtab}
\usepackage{multirow}

\begin{document}

\title{Project Life Cycles in Open-Source Software}

\author{
\IEEEauthorblockN{Sanjiv R. Das, Andrii Ieroshenko, Piyush Jain, David Qiu, Michael Chin, Brian E. Granger}
\IEEEauthorblockA{\{sanjivda, aieroshe, pijain, dlq, chnmch, brgrange\}@amazon.com}
}

\maketitle

\begin{abstract}

Using methods previously applied to product life cycles, this paper models developer engagement through the project life cycle for open-source projects, and detects similar dynamics in a cross section of projects. Endogenous growth theory is used to model growth dynamics in open-source software engineering, while incorporating the interactions between growth levels and developer activity over time using  systems of differential equations. The solution to this model calibrates well to many open-source projects. The model generates an estimate of the lifetime developer engagement and growth, which supports estimating a lifetime production value of open-source projects. 

\end{abstract}

\begin{IEEEkeywords}
open-source; growth, developers, endogenous growth theory
\end{IEEEkeywords}

\section{Introduction}

Changes in productivity in modern economies have been greatly ascribed to technological change, prompting the modeling of the path of technological advancement. In this paper, we model how growth levels (in open-source software) dynamically interact with the number of developers in the project. This interaction is important---as an open-source software (OSS) project grows and finds interest in the user community, a developer community also springs up around it. This accelerates the activity level of the project, which we measure as the total commits (additions and deletions) of lines of code in the software repository housed on platforms such as GitHub. As activity increases, it attracts more developers who wish to tweak and enhance the open-source project for their own use. This flywheel persists until the project matures and the level of activity slows down and the number of developers stabilizes (and often declines). These dynamics are similar to those considered in endogenous growth theory developed in \cite{romer_endogenous_1990}. A management framework for this interaction between developers and open-source project growth is highlighted in \cite{dey_cross_2024}---their CROSS model provides a foundation for understanding and enhancing the sustainability of OSS projects with a view to  practical application.

The term ``open source'' was coined on February 3, 1998, see  \cite{peterson_how_2018}. Open-source software (OSS) has achieved amazing strides and its role in the economy is undervalued and uncounted in GDP as it is free. The Synopsis Report \citep{bals_2024_2024} calls OSS everything, everywhere, all at once. It notes that the percentage of commercial code bases containing OSS is huge, on average 96\%. And 84\% of codebases contain at least one open-source vulnerability. 

There are two views on the value of open-source software. (i) The supply-side view that looks at the cost to create OSS, also known as the cost-based valuation. (ii) A demand-side view, i.e., the value created for the users of OSS, who do not have to build the software for themselves or pay for it. \cite{hoffmann_value_2024} estimate the supply-side value to be \$4.15 BN, and the demand-side value to be \$8.8 TN (which says that demand-side value is 2120x of supply-side value!). Developer time and cost can be measured from activity on OS repositories, but time and cost are extremely variable. Imputing labor costs to rewrite existing OSS arrives at estimates of \$38 BN in 2019 for the US \citep{robbins_first_2021} and Euro 1 BN in 2021 \citep{blind_impact_2021, blind_estimating_2024}. Assessments of value can be highly user-specific, and an interesting study reported in \cite{chesbrough_measuring_2023} brings out the various features and value of usage of OSS, noting that most firms find that cost savings and faster development are the main drivers of usage\footnote{\url{https://www.prnewswire.com/news-releases/linux-foundation-research-shows-economic-value-of-open-source-software\\-rising-in-terms-of-benefits-vs-costs-301761106.html}} \citep{wladawsky-berger_impact_2022, wladawsky-berger_whats_2024}.

We examine each commit to any chosen library over time and count the number of edited lines of code (added and removed) as a measure of technology progress in the repository. To illustrate, take the example of a popular open-source library, {\tt pandas}, released in 2009. {\tt pandas} is a powerful Python library for data manipulation and analysis. It offers fast and flexible data structures, easy handling of labeled and relational data, a high-level tool for practical data analysis in Python, with an ambitious goal of becoming the most comprehensive data analysis tool across programming languages. The library is designed to make data processing intuitive and efficient, serving as a crucial component for data scientists and analysts working with Python. 

\begin{figure}
\centering
\includegraphics[scale=0.65]{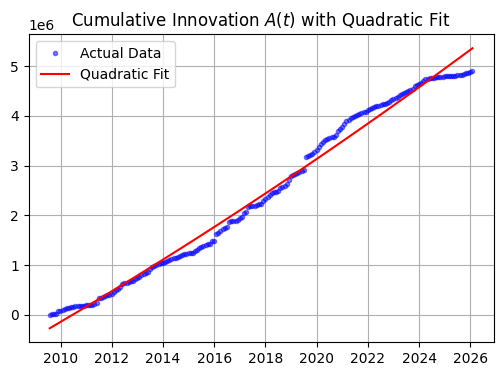}
\includegraphics[scale=0.65]{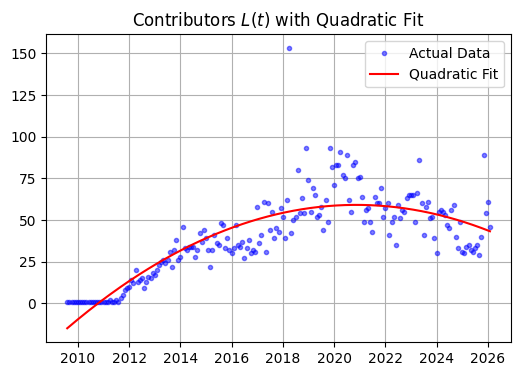}
\caption{\label{A_t} \small Plot of the cumulative lines of code changed (top plot) and the number of contributors (bottom plot) since the inception of the {\tt pandas} library in 2009. Lines added and deleted are counted as work done on the repository. The dots represent the effect of each commit. A second-order polynomial trend line was fitted to the code changes data, resulting in the following best-fit equation: $A(t) = 1.79 \times 10^{-2}t² + 8.26 \times 10^{2}t -2.71 \times 10^{5}$, $R^2 = 0.9879$. The fit of the number of contributors over time is represented as the following best fit polynomial trend of order two: $L(t) = -4.35 \times 10^{-6} t^2 + 3.59 \times 10^{-2} t - 1.49 \times 10$, $R^2 = 0.6968$. }
\end{figure}

Counting the changes over time to the repository, the cumulative growth in the project, $A(t)$, is shown in the top panel in Figure \ref{A_t}, and the number of contributors $L(t)$ per month is shown in the bottom plot. The models in this paper calibrate dynamical equations that represent these growth and contributor patterns, which may then be used to model the future path of any repository. 

These dynamical models may be relevant in answering many questions: (1) Can we understand developer engagement and is there are pattern across open-source projects? (Yes, the evidence shows the same S-curve as seen with product lifecycles). (2) How does developer engagement interact with project growth? (We find that some projects’ growth elasticity with developer engagement is positive and for others it is negative, leading to two different trajectories). (3) Is project growth endogenous and conforming to a version of \cite{romer_endogenous_1990}'s growth model? (Yes). (4) Can we use the models for developer engagement and growth to project time to maturation of the project when it does not need additional resources than just a developer? (Yes). (5) Can this lifecycle information help in judging lifetime OS project value? (Yes, though it depends on valuation assumptions). (6) Is there a constant ratio of project success (downloads) to developer effort? (No, there is wide cross-sectional variation in project success relative to developer). 

The rest of the paper proceeds as follows. In Section \ref{sec:endogenous_growth_model}, we describe an initial  dynamical equation used to model lines of code committed (added and deleted) to a repository. Section \ref{sec:dev_engagement} models the life cycle of developer engagement, which feeds into project growth. Section \ref{sec:calibration} calibrates the model to actual data for popular open-source projects. It also shows how to make lifetime projections and assesses stability of these forecasts at different stages in the project's life. Section \ref{sec:valuation} makes speculative suggestions about how the information generated from analyzing the project life cycle may be used to arrive at lifetime value. And Section \ref{sec:conclusions} provides closing comments.

\section{The Endogenous Growth Model}
\label{sec:endogenous_growth_model}

This model presumes that technological progress is not exogenous but driven by intentional investment decisions of agents (in our case, the developer ecosystem). Technology is treated as a non-rival good. Non-rival means that it can be used by multiple actors simultaneously. Further, in the open-source setting, goods are non-excludable, i.e., the owner does not exclude others from using it. In the open-source world the community owns the software and engagement is seamless. Most technologies tend to be excludable and often rival, which sets open-source software apart from other avenues of technological progress. We use these ideas to present a framework for understanding how decentralized growth can drive technological progress.  \cite{romer_origins_1994} offers a good review of endogenous growth models.

We introduce an adaptation of an endogenous growth model here with a basic model for developer engagement, which is refined later in Section \ref{sec:dev_engagement}. 
As noted previously, we define cumulative growth over time as variable $A(t)$ and developers (contributors) as labor $L(t)$. We use the Cobb-Douglas production functional form for changes in project growth.\footnote{\url{https://en.wikipedia.org/wiki/Cobb–Douglas_production_function}} This version of the model defines the rate of change in OSS growth as a function of current growth levels and the number of developers working on the project. We assume that the rate of change in developers is a simple function of the current engagement in the project (the model in the ensuing sections is more elaborate and will be used for the empirical work in this paper, but here, we introduce ideas with a much simpler model for developer engagement). With this assumption, the model admits a closed-form solution, shown below. (Later, we will enhance the developer model to include an independent propensity to contribute.)

Let the dynamics of growth level $A$ be described by the following function of labor $L$ (contributors). Changes in $A$ are also a function of the level of $A$ itself. This is what is known as the ``standing on the shoulders of giants'' effect (Isaac Newton, 1675), i.e., previous technological advancements drive new ones.     
\begin{equation}
{\dot A} = \frac{dA}{dt} = \gamma L^{\lambda} A^{\phi} \label{dotA}    
\end{equation}
where 
\begin{itemize}
\item $\gamma>0$ is a constant ``efficiency'' parameter;
\item $\lambda$ is elasticity of labor; if $\lambda=1$ doubling $L$ doubles the growth rate. If $\lambda<1$, there are diminishing returns to adding more contributors. If $\lambda>1$, there are increasing returns to adding contributors.
\item $\phi$ is growth spillover. If $\phi<0$, growth makes new growth harder (the ``fishing out'' effect). If $\phi=0$, new growth is not a function of past growth. If $0<\phi<1$, there are positive but diminishing returns to existing growth. If $\phi=1$, new growth is linear in existing growth. And if $\phi>0$, growth is explosive.
\item Scale effects: If $\lambda+\phi>1$, there are increasing returns to scale. If $\lambda+\phi=1$, there are constant returns to scale. If $\lambda+\phi<1$, there are decreasing returns to scale.
\end{itemize}
{\it Labor $L$}: A simplistic model for labor dynamics is the following:
\begin{equation}
{\dot L} = \frac{dL}{dt} = n L  \label{dotL}
\end{equation}
With initial conditions:
$$
L(0) = L_0, \quad A(0) = A_0
$$

The closed-form solution for both processes is (derivation details are in the Appendix):
{\small
\begin{eqnarray}
L(t) &=& L_0e^{nt} \label{Lsol}\\
A(t) &=& \left(\frac{(1-\phi)\gamma L_0^\lambda}{\lambda n}e^{\lambda nt} + A_0^{1-\phi} - \frac{(1-\phi)\gamma L_0^\lambda}{\lambda n}\right)^{\frac{1}{1-\phi}}  \label{Asol}
\end{eqnarray}
}
This solution is valid for $\phi \neq 1$, $n>0$. If $\phi = 1$, the equation would need to be solved differently. Also, for $n=0$, the solution is shown in equation (\ref{n0}). To check the solution by differentiation, set equation (\ref{Asol}) to $A(t)=u(t)^{1/(1-\phi)}$. so that $\frac{dA}{dt}=\frac{1}{1-\phi} \cdot u^{\frac{\phi}{1-\phi}} \cdot \frac{du}{dt}$, and then proceed to recover equation (\ref{dotA}) in a few steps of algebra. 

The overall solution shows that: (a) Labor grows exponentially at rate $n$ (this is by assumption, of course). (b) The growth level A grows as a function of both the initial conditions and the parameters $\gamma$, $\lambda$, $\phi$, and $n$.

This is a simple model where $L$ is modeled as an exponential growth process independent of $A$. As we can see, for example, in Figure \ref{A_t}, the number of contributors (labor level $L$) does not grow exponentially forever, so the ordinary differential equation for $L$ will need modification. We extend the model for $L$ in the next subsection, and this will also admit a closed-form solution for the model of developer engagement.  

\section{An engagement model for the number of developers}
\label{sec:dev_engagement}

We model the number of developers per month engaging with a project  using the \cite{bass_new_1969} and \cite{bass_why_1994} models, which are used for product life cycles. This approach has so far not been applied to software project life cycles. The main idea of the model is that the developer engagement rate per month of a project comes from two sources: 
\begin{enumerate}
\item The propensity of developers to work on the software project in any month because it is apt for their use case or from other motivations, independent of the other contributors propensity to engage in any month.
\item The additional propensity to work on the software in any month because others have adopted or are engaged with the project. Hence, at some point in the life cycle of a good open-source software project, social contagion, i.e., the influence of the early contributors becomes sufficiently strong so as to drive many others to work on developing the software as well. This is a network effect.
\end{enumerate}

We note that this definition of engagement deviates from the original specification in product adaption models where adoption is assumed to happen just once in a person's lifetime, whereas in our model, we are more interested in modeling repeated engagement of any developer. Therefore, we define engagement (or re-engagement) propensity {\em per month} using the model described below. 

The unit of developer engagement in an open-source project is taken to be a ``developer-month'' -- and the total number of developer-months over the lifetime of the project is denoted as $m$, which is latent, i.e., unknown until the project ends. Note also that $m = \sum_{t=1}^T L(t)$. Once a project is underway and has seen some traction, engagement statistics may be used to develop a forecast of future engagement. This will enable an estimation of total engagement $m$ and its distribution over the life cycle of the project. 

Define the fraction of a developer engagement in an OS project in month $t$ as $f(t)$. Hence, the number of developers engaging in month $t$ is simply $L(t) = f(t) \cdot m$. (Note that this implies $\int f(t) dt = 1$, and we may think of $f(t)$ as a probability.) So, the probability of engagement at time $t$ is the density function $f(t)=F'(t)$, where $F(t)$ is the cumulative probability or fraction of engagement up to time $t$. We can re-express this as a rate, conditional on remaining total engagement thus far, i.e., $\frac{f(t)}{1-F(t)}$. A functional form for this rate is the diffusion equation
\begin{equation}
\frac{f(t)}{1-F(t)} = p + q \cdot F(t) \label{diffusion}
\end{equation}
where $p$ is the independent rate of engagement (the coefficient of independent engagement) and $q$ modulates the network effect (the coefficient of imitation). The idea is that engagement is driven by a developer’s independent need for the software package and also driven by awareness from engagement by other developers. We may rewrite this equation as:
\begin{equation}
f(t) = [p + q \cdot F(t)][1-F(t)]
\end{equation}
The right-hand side shows that the fraction $f(t)$ may be expressed as the product of two components; the first $p+q\;F(t)$  may be interpreted as the current intensity of engagement, and the second is the remaining engagement fraction.   

The goal here is to fit this equation to software developer engagement data to ascertain $p,q$ and the lifetime engagement in developer months (denoted $m$). We rewrite the equation above as 
\begin{equation}
\frac{dF/dt}{1-F} = p + q \cdot F, \quad F(0)=0
\end{equation}
which is a differential equation with an initial condition, which we can solve for $F(t)$, provided here: 
\begin{equation}
F(t) = \frac{p[e^{(p+q)t}-1]}{p e^{(p+q)t}+q}
\end{equation}
Differentiating, we get the engagement probability density at time t:
\begin{equation}
f(t) = \frac{dF}{dt} = \frac{e^{(p+q)t} p (p+q)^2}{[p e^{(p+q)t}+q]^2} \label{ft}
\end{equation}

Denoting the lifetime engagement as $m$, then expected engagement at time $t$ is $L(t)=m \cdot f(t)$. Using data, we want to estimate $\{p,q,m\}$. Cumulative engagement up to time $t$ is defined as ${\cal L}(t)=m \cdot F(t)$. Substituting these into the diffusion equation (\ref{diffusion}) above we have 
\begin{equation}
\frac{L(t)/m}{1-{\cal L}(t)/m} = p + q \cdot U(t)/m
\end{equation}
Re-arranging we have
\begin{eqnarray}
L(t) &=& [p+q \; {\cal L}(t)/m][m-{\cal L}(t)] \nonumber \\
 &=& \beta_0 + \beta_1 {\cal L}(t) + \beta_2 {\cal L}(t)^2 \\
\beta_0 &=& p\; m \\
\beta_1 &=& q-p \\
\beta_2 &=& -q/m
\end{eqnarray}
If we have data on periodic engagement $L(t)$ and cumulative engagement ${\cal L}(t)$ we can fit a regression to the data to get coefficients $\beta_0, \beta_1, \beta_2$. We can then solve for $p,q,m$ as follows. Given that
\begin{equation}
\beta_1 = q-p = -m\; \beta_2 - \beta_0/m,
\end{equation}
we re-arrange to get a quadratic equation
\begin{equation}
\beta_2 m^2 + \beta_1 m + \beta_0 = 0
\end{equation}
with solution for lifetime engagement $m$:
\begin{equation}
m = \frac{-\beta_1 \pm \sqrt{\beta_1^2 - 4 \beta_0 \beta_2}}{2 \beta_2}
\end{equation}
The positive root of $m$ may then be used to solve for 
\begin{equation}
p = \beta_0/m,  \quad \quad q = -m \beta_2
\end{equation}

Next, we model how growth (lines of code) changes dynamically over time. Models of growth have been used to value open-source projects \citep{horowitz_software_2001, hoffmann_value_2024} and obtaining lifetime value begins with estimating lifetime growth. 

\subsection{Rescaling}
\label{sec:rescaling}

If we rescale time by $t_0 = \frac{1}{p+q} \ln(q/p)$ and $f$ by $f_0 = \frac{(p+q)^2}{4q}$ (these are the time and height of peak developer engagement) we get normalized developer engagement:
$$
f' = f/f_0 = \mbox{sech}^2 \left( \frac{\alpha}{2} (1 - t') \right)
$$
where $\alpha = ln(q/p)$ and $t' = t/t_0$. Developer engagement is scaled and shifted by the function $\mbox{sech}(\cdot)^2$. Since $\alpha$ indicates that just the ratio $q/p$ matters, this is effectively a single parameter model, modulated by the ratio of the coefficients of the network effect and the independent effect. 

\section{Calibrating Developer Engagement and Growth to Data}
\label{sec:calibration}

We download all commits in a project from GitHub and count the number of lines added and deleted in each commit, from the inception of the project. We use the contributor information from these commits to count the number of unique developers contributing to a project in every month. First, we fit the data to the model for developer engagement. Second, we fit the dynamics of the growth model to the data on commits and the data on developer engagement. The code for this project is available here: \url{https://github.com/srdas/oss-lifecycle}. 

\subsection{Fitting developer data}

\begin{figure}
\centering
\includegraphics[scale=0.6]{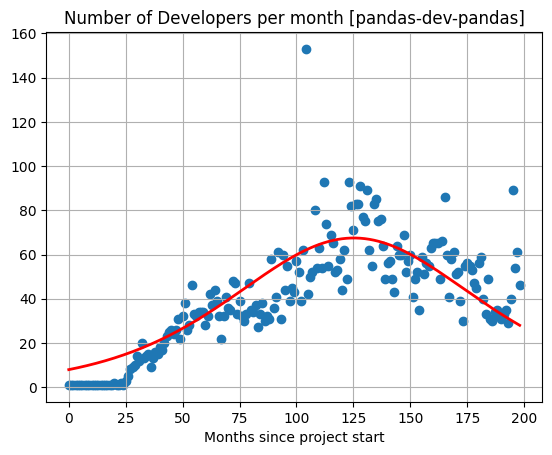}
\caption{\label{engagement} \small Plot of the developer engagement  since the inception of the {\tt pandas} library in 2009. The fitted line uses the solution in the equations above where it is determined that $p=0.00084$, $q=0.02686$, and $m=9448$. }
\end{figure}

As an example, the model for developer engagement is fit to the data for the {\tt pandas} library using the closed-form solution above. Figure \ref{engagement} shows the monthly engagement data, and the bottom panel shows the engagement rate. The best-fit parameters are $p=0.00084$, $q=0.02686$, and $m=9448$. This calibration is straightforward because of the availability of the closed-form solution in Section \ref{sec:dev_engagement}. Estimates for a collection of popular open-source projects are presented as examples in Table \ref{os_developer_model_stats}. We also note here that the parameters $p,q$ have the same scale as those in the original \cite{bass_new_1969} paper (see Table I there).

\subsection{Fitting growth data}

The behavior of the endogenous growth system will be a function of the three growth parameters $\{\gamma, \lambda, \phi\}$ in equation (\ref{dotA}), which is an ordinary differential equation (ODE). Using the solution to the ODE with a given set of these parameters enables tracing out the estimated function ${\hat A}(t)$ over time using the data for both, commits and developer engagement. While $\gamma$ is a scaling constant the parameters $\lambda, \phi$ are elasticities with respect to developer engagement and project growth levels. The sign of these parameters informs us about how the project responds to infusions of each resource. 

For calibration, we minimize the root mean squared error between the data and estimated functions from solving the ODEs. The calibration objective function is:
\begin{equation}
\min_{\{\gamma, \lambda, \phi\}} \frac{1}{T}\sum_{t=1}^T \left({\hat A}(t, {\hat L}(t))-A(t) \right)^2 
\end{equation}
where $T$ is the number of months in the data and ${\hat A}(t, {\hat L}(t))$ comes from the solution to the ODE for $A$ and uses the solution for the ODE for $L$ to estimate ${\hat L}(t)$.

\begin{figure}
\centering
\includegraphics[scale=0.6]{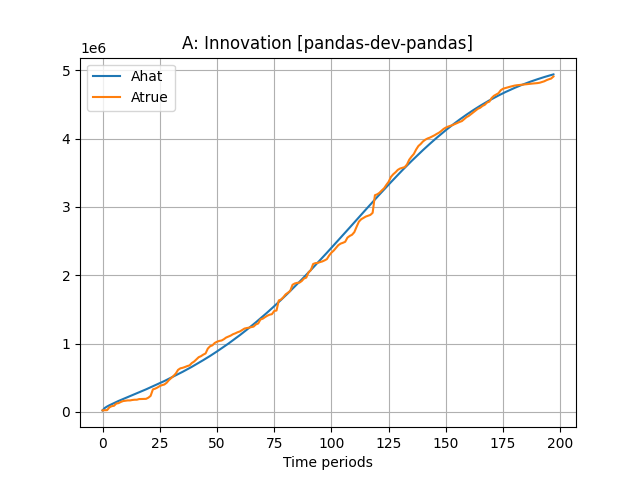}
\caption{\label{growth} \small For the {\tt pandas} library, the plot shows the fit against the raw data after calibrating the ODE of the Cobb-Douglas model to the data. Best-fit parameters: $\gamma = 601657.05$, $\lambda = 1.301$, $\phi = -0.552$. Time periods are months. 
}
\end{figure}

Numerically, each iteration of the minimizer calls the ODE solver internally to generate an estimated dynamical path for growth ${\hat A}$, which is then compared to the true path $A$ by the optimizer. The ODE parameters are updated and the objective function is refined until parameters are found that minimize it. The solution is obtained in a few seconds and converges to the minimum. Once the optimal parameters are determined, these are used to plot the estimated functions against the actual commits data to assess the fit. For the {\tt pandas} library, see Figure \ref{growth}. We see a very good fit. Labor elasticity ($\lambda$) is positive and growth elasticity ($\phi$) is negative. Adding more contributors brings more growth, but growth eventually saturates as seen from the negative elasticity for growth. 

\begin{table} 
\centering
\caption{\label{tab:innov_params} \small Calibrated parameters for the growth model, $\frac{dA}{dt} = A \cdot L^{\lambda} \cdot A^{\phi}$. $\gamma$ is a constant, $\lambda$ is the elasticity of growth with respect to developer engagement, and $\phi$ is the elasticity of growth with respect to current growth. }
\begin{tabular}{lccc}
\toprule
Project & $\gamma$ & $\lambda$ & $\phi$ \\
\midrule
dask-dask & 13255.95 & -0.3811 & 0.0111 \\
huggingface-transformers & 292.79 & -0.4171 & 0.5017 \\
jupyterlab-jupyterlab & 11300.14 & 1.4304 & -0.2231 \\
kubernetes-kubernetes & 0.01 & 12.9150 & -1.1730 \\
langchain-ai-langchain & 1431.86 & 0.7093 & 0.0916 \\
microsoft-DeepSpeed & 649.79 & -0.6623 & 0.3490 \\
numpy-numpy & 370.42 & -0.3206 & 0.3318 \\
pandas-dev-pandas & 601657.05 & 1.3005 & -0.5523 \\
pytorch-pytorch & 21.03 & 0.4524 & 0.4228 \\
vllm-project-vllm & 13.30 & 0.1723 & 0.5940 \\
\bottomrule
\end{tabular}
\end{table}

For ten example projects shown in Table \ref{tab:innov_params} we see that a majority of the projects have a positive elasticity with developer engagement, where $\lambda>0$ and also a positive one with growth level, $\phi>0$. This means that, over time, projects get a stable pool of contributors who improve their skills in the project and are able to accelerate growth ($\lambda>0)$. This is good because over time, developer engagement peaks and then declines. Because $\phi>0$, as the project matures and cumulative growth level $A$ grows, the pace at which lines of code are added also increases. A similar pattern has been noticed by venture capitalists \citep{droesch_measuring_2020}. 

\subsection{Extrapolation}

\begin{figure*}
\centering
\includegraphics[scale=0.45]{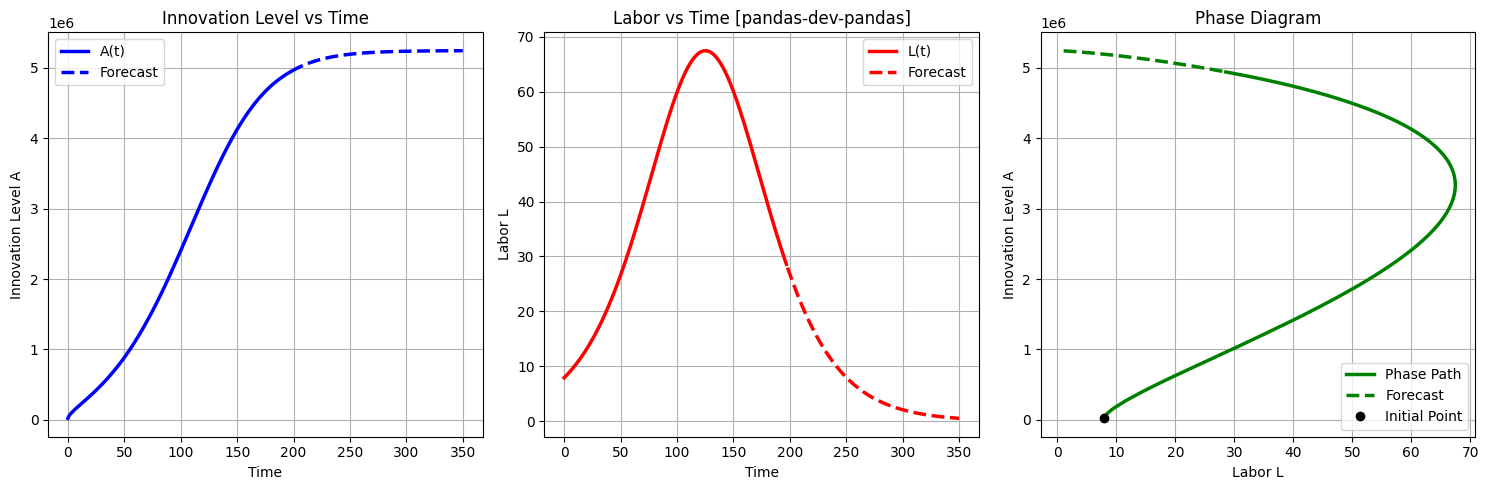}
\caption{\label{ALprojections} \small Extrapolation of the solution for the {\tt pandas} repository till its maturation, which is defined as engagement dropping to half a developer per month. The phase diagram shows the interaction between $L$ and $A$, and we can see that as the project matures, growth grows without the need for a large number of additional contributors.  }
\end{figure*}

In order to assess the future trajectory of the project the ODE solutions may be projected further for additional months. The three projection plots for the {\tt pandas} project are shown in Figure \ref{ALprojections}. The phase diagram shows the interaction between $L$ and $A$, and we can see that as the project matures, growth continues without the need for a large number of additional contributors, consistent with the increasing returns and the long-run growth model in \cite{romer_increasing_1986}. 

We can also ask when developer engagement will taper off and stabilize, as it inevitably must for all projects. Bessemer Venures denoted this point as the ``steady-state maturity'' of the project \citep{droesch_measuring_2020}. We implement this by solving for the value of $t$ that sets $f(t) \cdot m = 0.5$ in equation (\ref{ft}). That is, engagement has dropped to a half-developer per month (no full time developer activity). Figure \ref{ALprojections} (middle plot) shows the start and end of developer engagement in addition to its life cycle forecast for the {\tt pandas} project. 

We conducted this analysis for other example projects and found the remaining life to maturation to be as follows, assuming no significant changes to the project: {\tt pandas} has 12.7 years remaining to maturation, and others, for example, are {\tt jupyterlab} 7.9 years, {\tt langchain} 1.82 years. Table \ref{os_developer_model_stats} shows the model parameters and life expectancy of each project. We see that the coefficients are similar to that obtained in work on product life cycles. More importantly, the $p$ and $q$ parameters are of similar size and scale across all projects, suggesting that the model applies well in the cross-section of open-source repositories. (For some projects $p$ or $q$ is negative, in which case the developer engagement data does not fit the model well.)

\begin{table*}
\centering
\caption{\label{os_developer_model_stats} \small Fitted parameters of the developer model in section \ref{sec:dev_engagement}. The start date and end date of the data used for each project is shown. The columns in the table are as follows: 1. $p$: coefficient of independent propensity (probability) to join the project in any month.
2. $q$: coefficient of network intensity (probability) to join the project in any month.
3. $m$: calibrated developer-months expected over the project's lifetime.
4. $t$: current life of project in months.
5. $T$: expected lifetime of project in months.
6. $yrs$: at the current rate, expected remaining time to maturation of a project in years after the End Date of the project.
All project fits showed high $R^2$ values with all t-statistics for parameters $p, q, m$ showing statistical significance at the 99\% level. 
}
\begin{tabular}{lcccccccc}
\toprule
Project & Start Date & End Date & $p$ & $q$ & $m$ & $t$ & $T$ & $yrs$ \\
\midrule
dask-dask & 2014-12-31 & 2026-01-31 & 0.00210 & 0.03855 & 2173.64627 & 134 & 199.97683 & 5.49807 \\
huggingface-transformers & 2018-10-31 & 2026-01-31 & 0.00230 & 0.03078 & 12940.35495 & 88 & 284.67205 & 16.38934 \\
ipython-ipython & 2005-07-31 & 2026-01-31 & 0.00149 & 0.02430 & 2817.00050 & 247 & 303.19720 & 4.68310 \\
jax-ml-jax & 2018-11-30 & 2026-01-31 & -0.01448 & 0.00058 & -1360.19200 & 87 & -258.61177 & -28.80098 \\
jupyter-server-jupyter-scheduler & 2022-09-30 & 2025-12-31 & 0.04236 & -0.06855 & 189.77015 & 40 & 47.53731 & 0.62811 \\
jupyterlab-jupyter-ai & 2023-02-28 & 2026-01-31 & 0.01421 & 0.11660 & 240.70389 & 36 & 48.41652 & 1.03471 \\
jupyterlab-jupyterlab & 2015-07-31 & 2026-01-31 & 0.00358 & 0.03007 & 2820.22037 & 127 & 222.22246 & 7.93520 \\
kubeflow-kubeflow & 2017-11-30 & 2026-01-31 & 0.02221 & 0.02332 & 897.26940 & 99 & 112.18981 & 1.09915 \\
kubernetes-kubernetes & 2014-06-30 & 2026-01-31 & 0.00464 & 0.02233 & 23872.21258 & 140 & 330.63036 & 15.88586 \\
kubernetes-sigs-kueue & 2021-11-30 & 2026-01-31 & 0.00191 & 0.04751 & 2854.25959 & 51 & 179.90429 & 10.74202 \\
langchain-ai-langchain & 2022-10-31 & 2026-01-31 & 0.01343 & 0.14676 & 5263.69827 & 40 & 61.84962 & 1.82080 \\
langchain-ai-langchain-aws & 2024-03-31 & 2026-01-31 & 0.00867 & 0.06620 & 540.68494 & 23 & 87.18043 & 5.34837 \\
langchain-ai-langchain-google & 2024-02-29 & 2026-01-31 & 0.02816 & 0.05942 & 499.36713 & 24 & 63.82810 & 3.31901 \\
microsoft-DeepSpeed & 2020-01-31 & 2026-01-31 & 0.00197 & 0.07003 & 1496.74400 & 73 & 124.48289 & 4.29024 \\
mlflow-mlflow & 2018-06-30 & 2026-01-31 & 0.00150 & 0.01044 & 11476.56955 & 92 & 643.39112 & 45.94926 \\
numpy-numpy & 2001-12-31 & 2026-01-31 & 0.00018 & 0.01735 & 9769.60912 & 290 & 593.24927 & 25.27077 \\
pandas-dev-pandas & 2009-07-31 & 2026-01-31 & 0.00084 & 0.02686 & 9448.61510 & 199 & 352.18475 & 12.76540 \\
pytorch-pytorch & 2012-01-31 & 2026-01-31 & 0.00064 & 0.02930 & 34554.29690 & 169 & 383.43603 & 17.86967 \\
ray-project-ray & 2016-02-29 & 2026-01-31 & 0.00126 & 0.02749 & 10234.99818 & 120 & 330.39570 & 17.53298 \\
run-llama-llama\_index & 2022-11-30 & 2026-01-31 & 0.00770 & 0.13915 & 3014.88442 & 39 & 66.27020 & 2.27252 \\
scikit-learn-scikit-learn & 2010-01-31 & 2026-01-31 & 0.00163 & 0.01677 & 10026.76782 & 193 & 452.60215 & 21.63351 \\
tensorflow-tensorflow & 2015-11-30 & 2026-01-31 & 0.00387 & 0.02769 & 28138.49944 & 123 & 303.58833 & 15.04903 \\
vllm-project-vllm & 2023-02-28 & 2026-01-31 & 0.00121 & 0.11101 & 12331.92116 & 36 & 110.96520 & 6.24710 \\
\bottomrule
\end{tabular}
\end{table*}

\begin{figure*}
\centering
\includegraphics[scale=0.56]{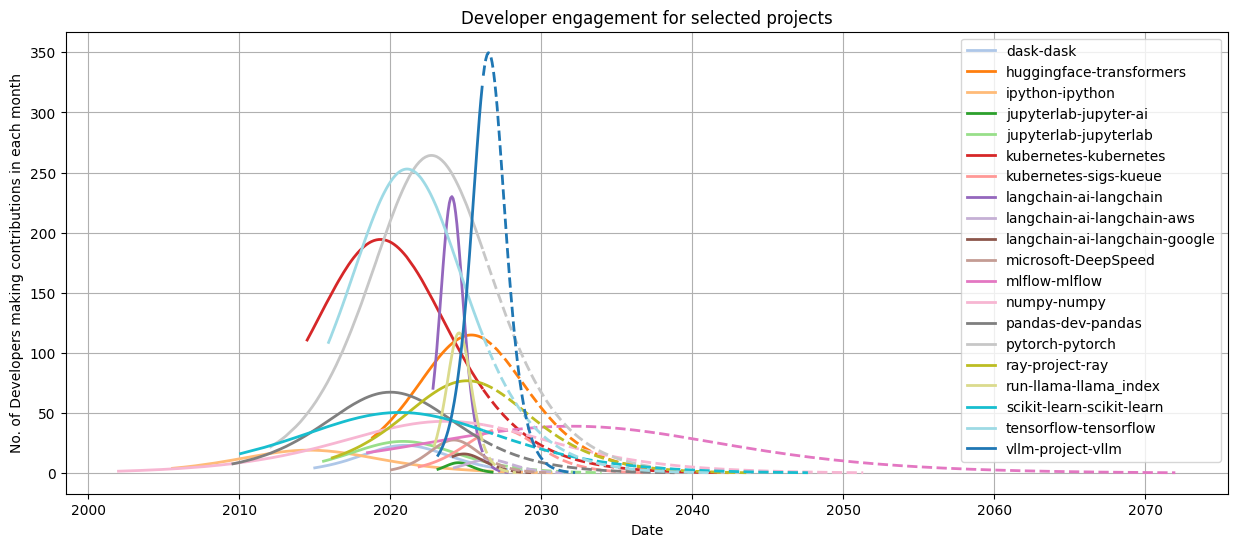}
\includegraphics[scale=0.56]{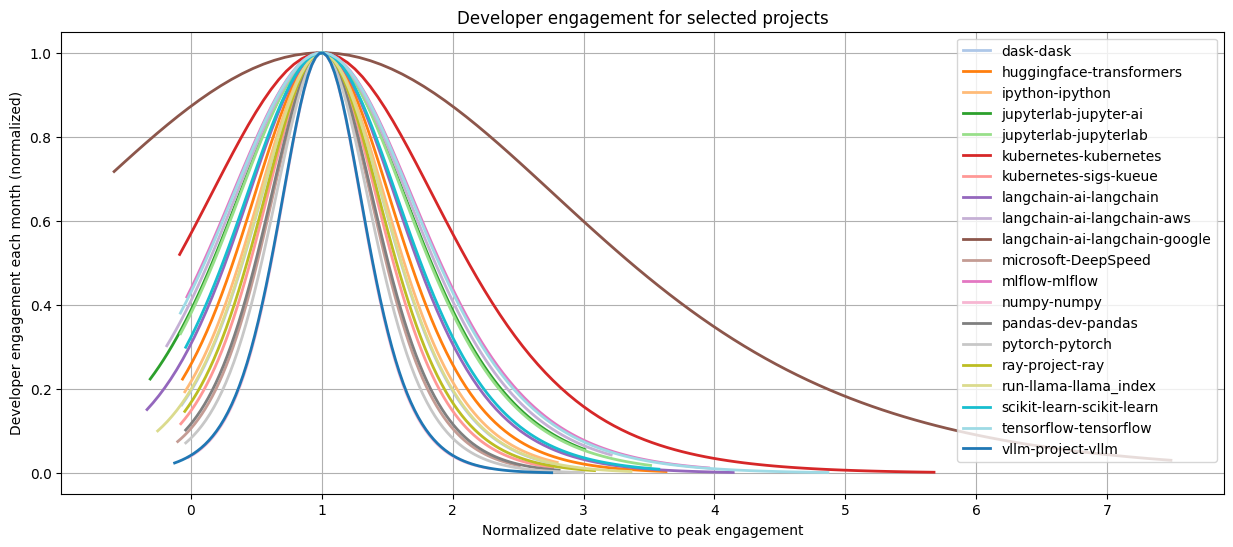}
\caption{\label{fig:all_lifecycles} \small Developer engagement on open-source projects over time, see also Table \ref{os_developer_model_stats}. Engagement is measured as the number of developers who commit code each month. The engagement plots have a dashed line that starts end January 2025 and shows the projected future trajectory of developers engaging until the maturation of the project. The bottom plot shows the same data normalized using the mathematical results in Section \ref{sec:rescaling}. The x-axis is normalized to show the time of peak developer engagement to be 1, and the y-axis is normalized so that the peak height is 1 for each project. Projects with greater overall lifespan are wider than others with lesser lifespans. 
}
\end{figure*}

\subsection{Forecast Stability}

\begin{figure*}
\centering
\includegraphics[scale=0.35]{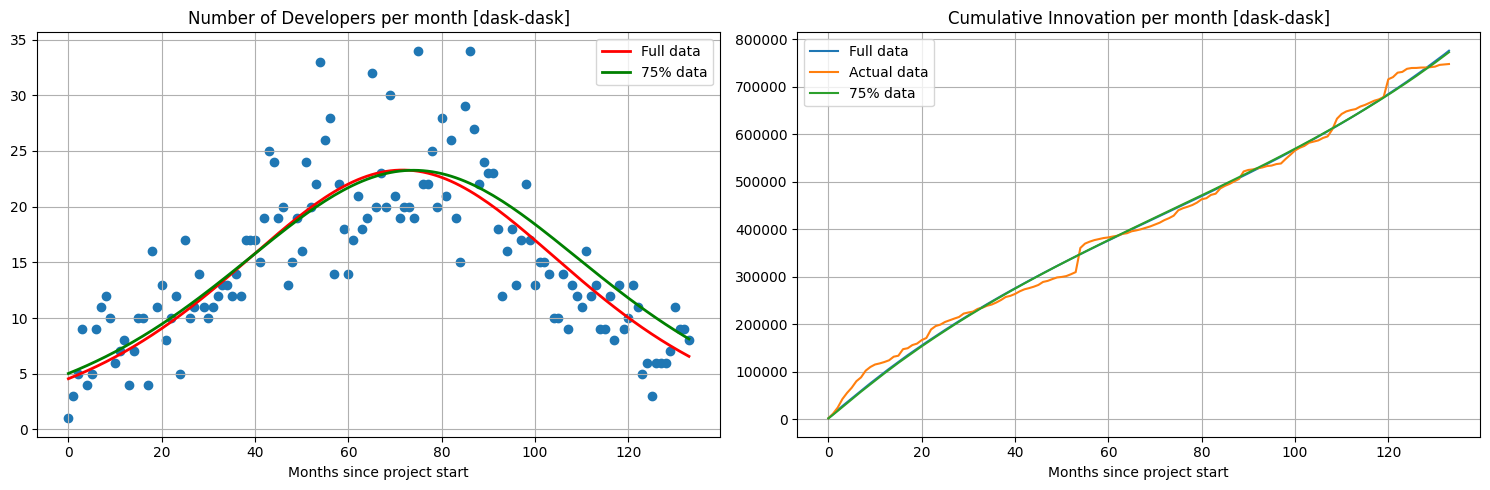}
\includegraphics[scale=0.35]{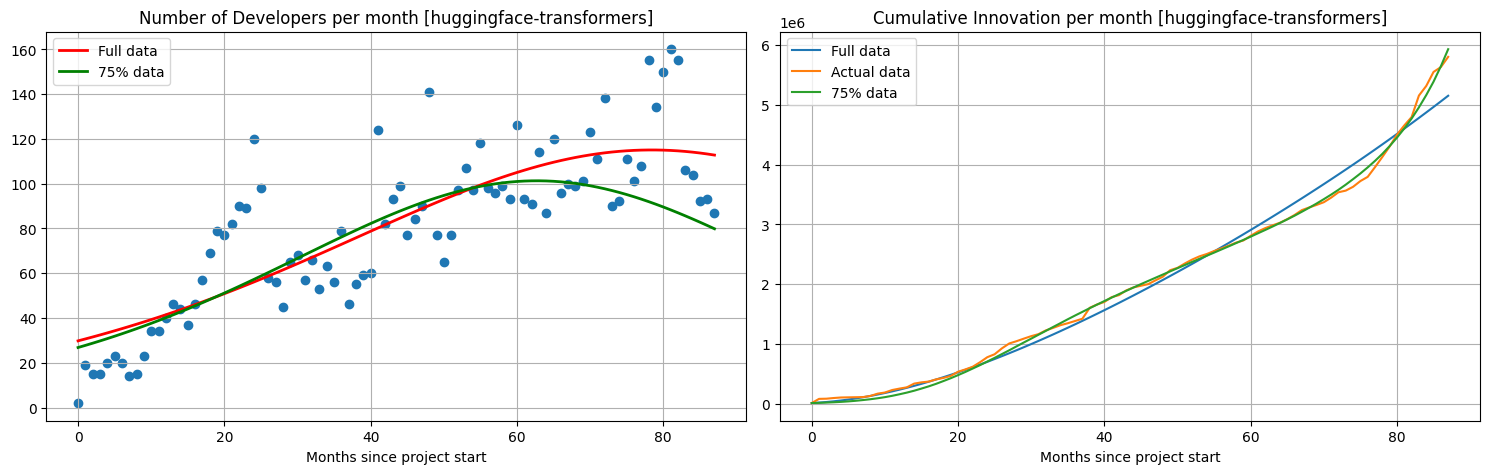}
\includegraphics[scale=0.35]{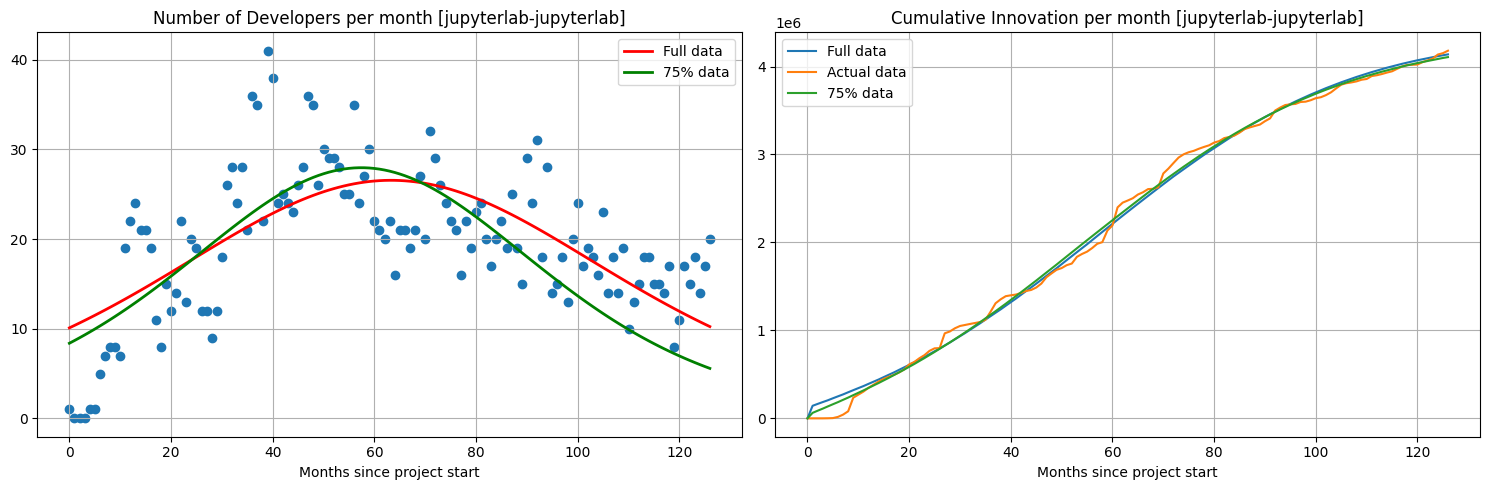}
\includegraphics[scale=0.35]{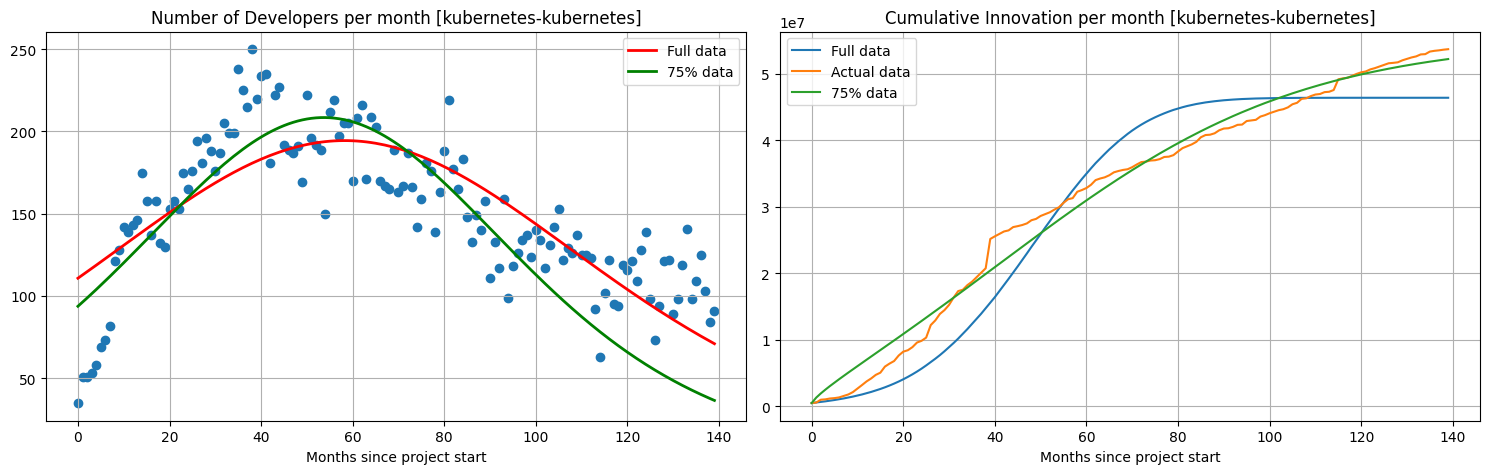}
\includegraphics[scale=0.35]{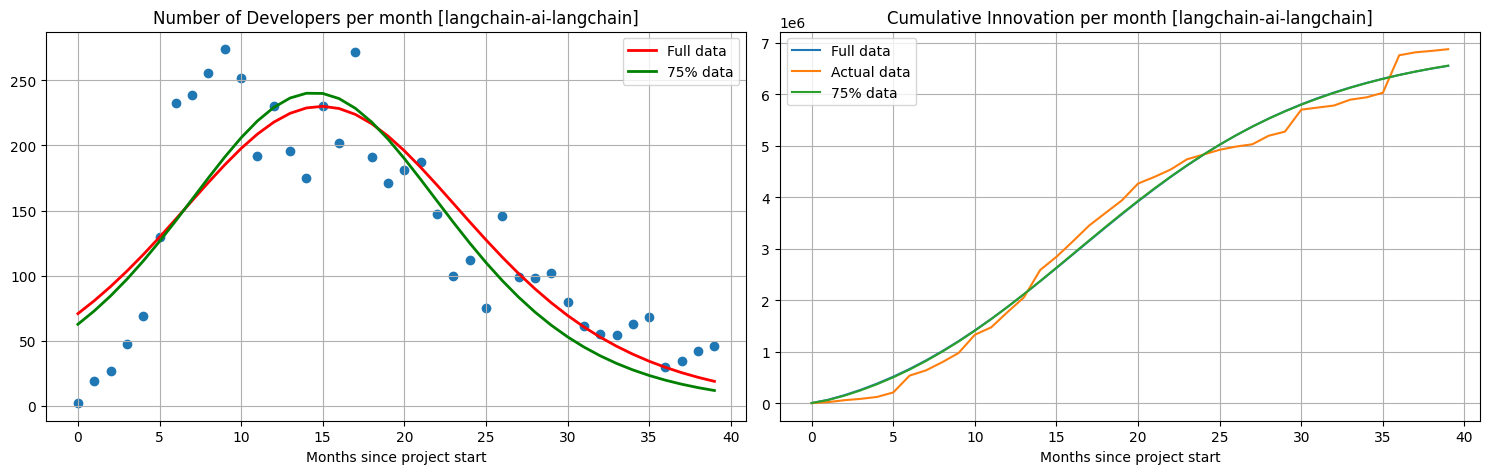}
\caption{\label{fig:model_fits1} \small Model calibration I: The plots show the fitted values to developer engagement data (left) and cumulative growth (right). The actual data is shown alongside the fitted data when the fit is undertaken using both, the entire data sample and the first 75\% of the data available for each project. }
\end{figure*}

\begin{figure*}
\centering
\includegraphics[scale=0.35]{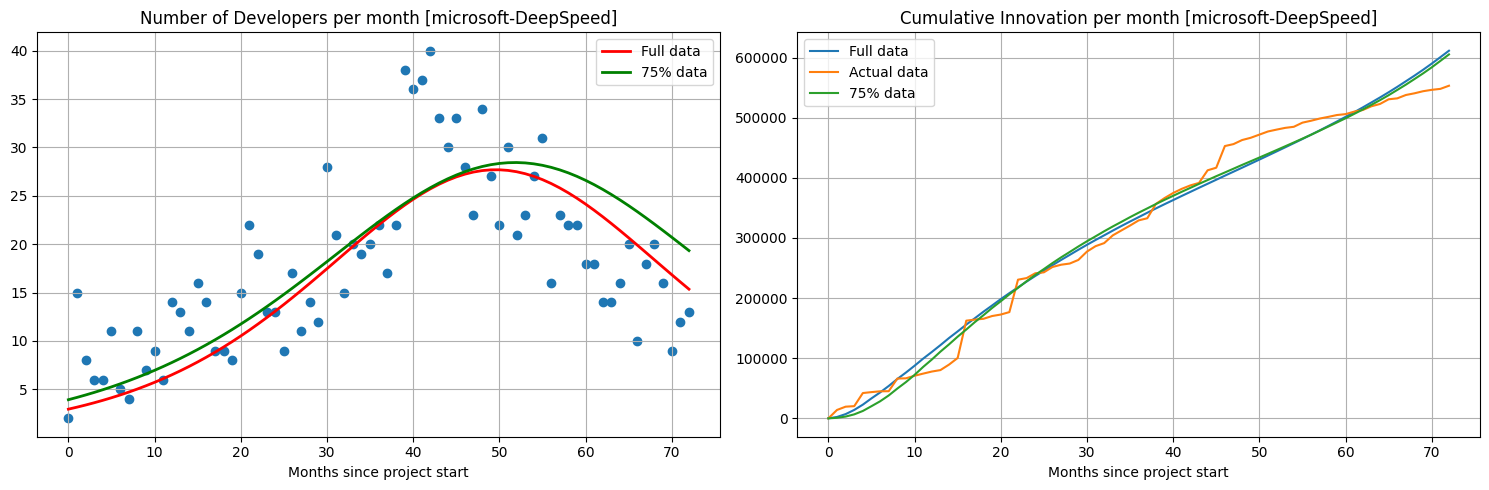}
\includegraphics[scale=0.35]{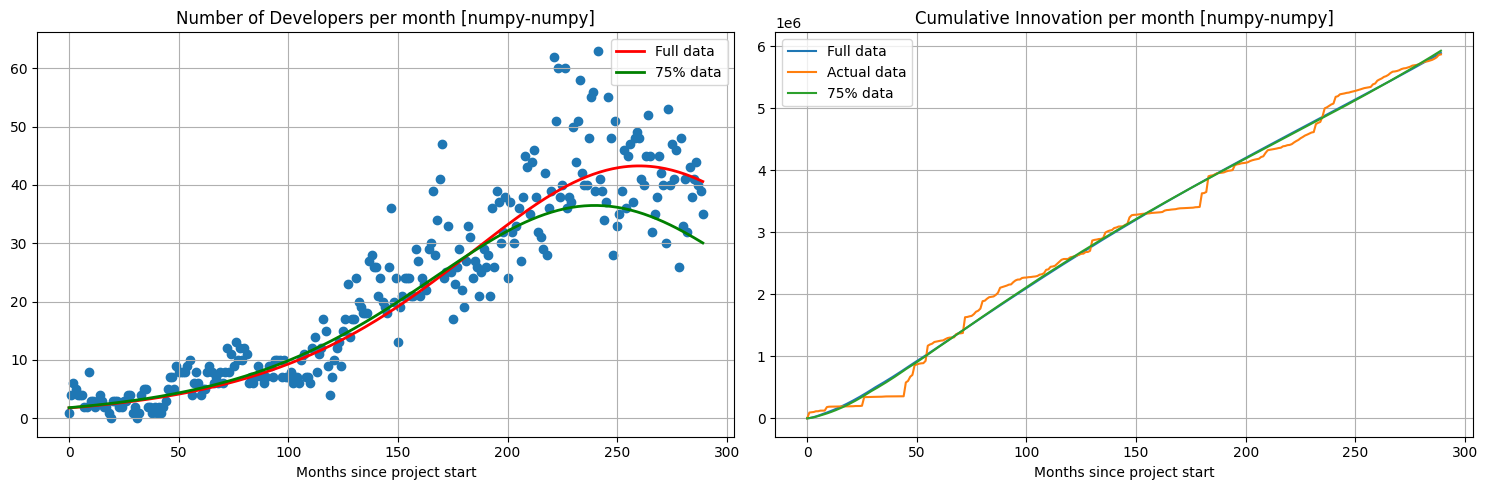}
\includegraphics[scale=0.35]{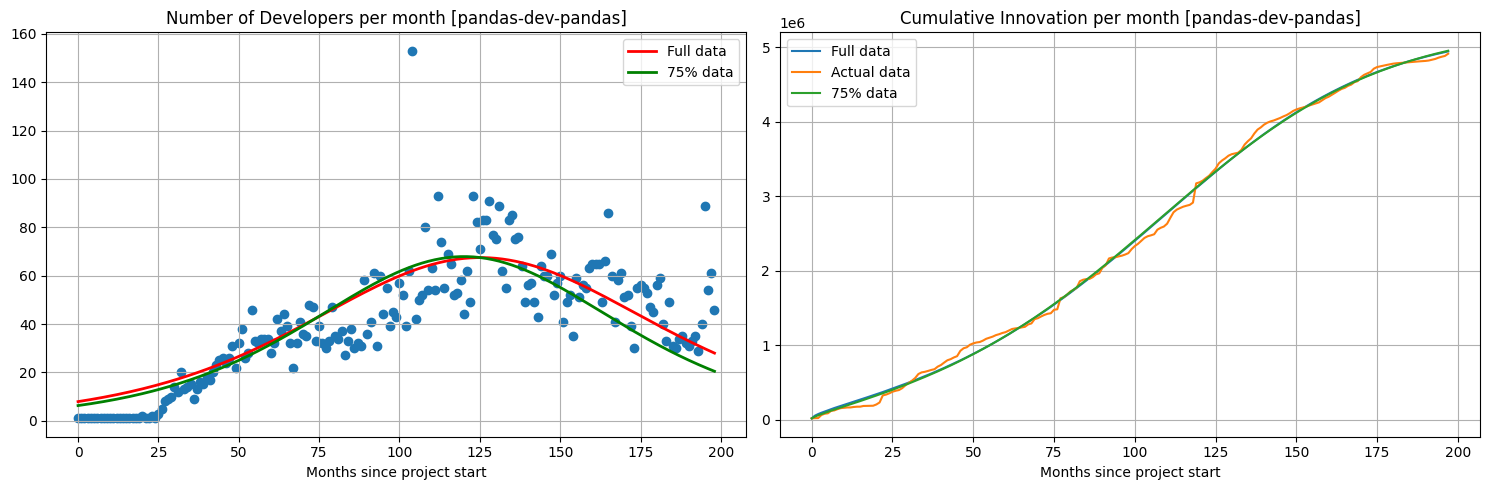}
\includegraphics[scale=0.35]{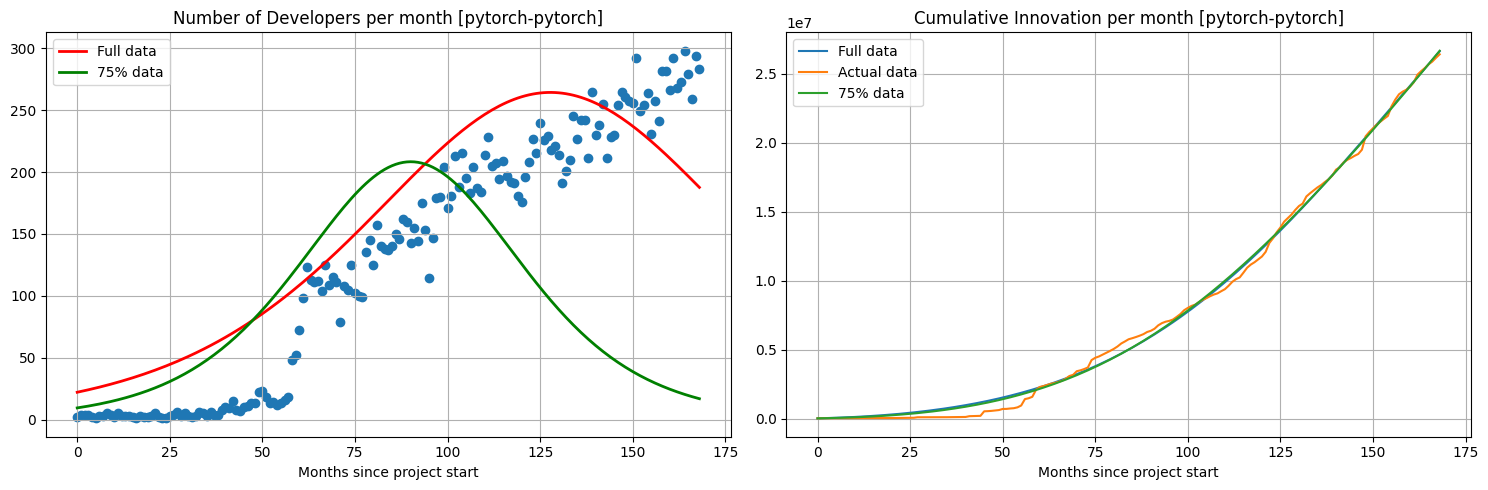}
\includegraphics[scale=0.35]{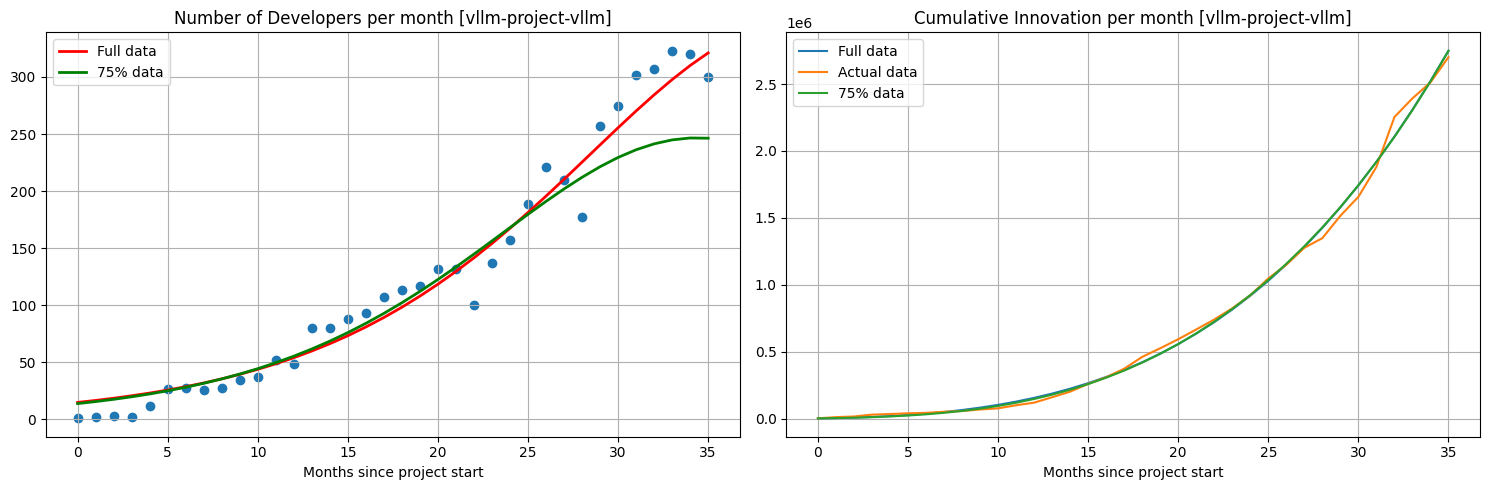}
\caption{\label{fig:model_fits2} \small Model calibration II: The plots show the fitted values to developer engagement data (left) and cumulative growth (right). The actual data is shown alongside the fitted data when the fit is undertaken using both, the entire data sample and the first 75\% of the data available for each project. }
\end{figure*}

Since the model may be calibrated to data at different points in the life of the project, and then used to forecast the future life of the project till its demise, it is useful to examine how stable these forecasts are for the time of maturation. Of course, as the time moves on, we gather more information about a project, which will mean making adjustments to the forecast of both developer engagement and growth levels. In order to assess how these forecasts change, we examined ten well-known open-source projects as follows. First, we calibrated the model using all the data about the project from its inception until the current time (end January 2026 at the time of data collection). Second, we plotted the fitted developer engagement and fitted cumulative growth and compared these to the actual data. As shown in Figures \ref{fig:model_fits1} and \ref{fig:model_fits2}, the fit of both the developer engagement model and the cumulative growth model is very good across most projects. 

Third, we then used only data for the first 75\% of the project's current life. For a project like {\tt pandas}, which began in 2009, this would mean dropping the most recent four years to get estimates. In the third row of Figure \ref{fig:model_fits2}, we see that the fitted developer engagement model (left plot) is almost the same when calibration uses full data (red line) versus when only the first 75\% of the project's lifetime data is used (green line). When considering the cumulative growth fit (right plot) the two lines are almost identical. Examining both figures \ref{fig:model_fits1} and \ref{fig:model_fits2} reveals that the future projections for the project are robust to the timing of calibration. Even for projects with a short life so far, this does not  matter for estimating growth dynamics, as they seem to fit very well regardless of when the calibration is done. However, for estimating developer engagement, sometimes the calibrations do lead to different fitted dynamics as in the case of {\tt pytorch}, though for most of the other projects, even for developer engagement the calibrations at different times lead to similar developer engagement plots. Researchers may use these forecasts to support valuation of these projects, so the fact that robust estimates of lifetime dynamics are available early in a project's life is encouraging. 

In Section \ref{sec:valuation} we will use these dynamic models for developer engagement and growth to suggest methods for valuation of OSS projects, though we surmise that these would vary based on assumptions made by the valuer.

\subsection{Adaptation to Closed Source Software}

These interactions in the open-source ecosystem may be used to estimate headcount needed to develop closed-source software using a simpler model with only equation (\ref{dotA}), given that developer labor will be set to a constant $L$. The solution is 
\begin{equation}
A(t) = \left[(1-\phi)\left(\gamma L^{\lambda}\cdot t + \frac{A_0^{1-\phi}}{1-\phi} \right) \right]^{\frac{1}{1-\phi}} \label{n0}
\end{equation}
This equation can be calibrated to the data for project growth by modulating the parameters $\gamma$, $\lambda$, and $\phi$ for a fixed level of $L$ using the same solution approach from the previous subsection. This is simpler than modeling the dynamics of open-source software projects. Extensions to the model for how developers are added to the closed-source project would be internal to and vary by organization.

\section{Valuation Approaches}
\label{sec:valuation}

\begin{table*}
\centering
\caption{\label{tab:valuation} \small lifetime statistics and estimated valuations for ten key open-source projects. Project life (time to maturation) is measured in months. ``Dev Months'' is developer months. Growth is in millions of lines of code committed. The ``Innov/Dev Ratio'' is the cumulative lines of code committed (both additions and deletions) divided by the cumulative number of developer months up to current life. The project value is calculated as Growth divided by the Innov/Dev Ratio multiplied by the cost per month per developer, assumed to be \$10,000, multiplied by 0.5, assuming that open-source developers spend 50\% of their work time on any project. Valuation here relates to the production cost of the project, also known as its supply-side value. 
}
\begin{tabular}{lcccccccc}
\toprule
Project & Current & Full  & Current Cum & Current Cum & Lifetime & Innov/ & \multicolumn{2}{c}{\underline{Supply-side Valuation}} \\
 & Life  & Life  & Developer & Growth  & Growth  & Dev-Month & Current  & To Maturation  \\
 & (months) & (months) & Months & (MM) & (MM) & (Ratio) & (\$MM) & (\$MM) \\
\midrule
dask-dask & 134 & 200 & 1938 & 0.78 & 1.58 & 403.98 & 9.69 & 19.57 \\
huggingface-transformers & 88 & 285 & 6990 & 5.25 & 250.80 & 750.41 & 34.95 & 1671.09 \\
jupyterlab-jupyterlab & 127 & 222 & 2426 & 4.15 & 4.39 & 1710.05 & 12.13 & 12.83 \\
kubernetes-kubernetes & 140 & 331 & 20949 & 46.41 & 46.41 & 2215.58 & 104.75 & 104.75 \\
langchain-ai-langchain & 40 & 62 & 5168 & 6.60 & 6.97 & 1277.78 & 25.84 & 27.27 \\
microsoft-DeepSpeed & 73 & 124 & 1213 & 0.62 & 3.37 & 513.22 & 6.07 & 32.85 \\
numpy-numpy & 290 & 593 & 5939 & 5.91 & 21.57 & 995.03 & 29.70 & 108.40 \\
pandas-dev-pandas & 199 & 352 & 8214 & 4.96 & 5.24 & 603.77 & 41.07 & 43.42 \\
pytorch-pytorch & 169 & 383 & 26430 & 26.93 & 56.88 & 1018.82 & 132.15 & 279.15 \\
vllm-project-vllm & 36 & 111 & 4465 & 2.99 & 35.49 & 669.48 & 22.32 & 265.08 \\
\bottomrule
\end{tabular}
\end{table*}

Table \ref{tab:valuation} reports lifetime developer engagement and growth, as well as illustrative (under assumptions) supply-side costs for valuation of the ten projects portrayed in Figures \ref{fig:model_fits1} and \ref{fig:model_fits2}. The current life of the project and forecast full life are shown. The project life is assumed to ``mature'' when developer engagement drops to half a developer a month. For many projects this is shown in Figure \ref{fig:all_lifecycles} (top, the bottom plot shows the normalized version of the top plot). We note that the project does not stop being downloaded and used, just that developer maintenance becomes minimal and reaches a low steady-state. 

The number of developer months currently invested in the project is shown in Table \ref{tab:valuation}. This is the sum of the number of developers each month for all months. Current cumulative growth is the total number of lines of committed code (additions and deletions) until current time and lifetime growth as forecast by the dynamic model is shown as well. Growth per developer month in the table is a useful statistic that will be used for valuation of the projects and it varies widely across projects. This is calculated based on data up to current time. 

\subsection{Supply-side value}

The valuations shown in the last two columns of Table \ref{tab:valuation} are calculated as follows. Growth is divided by the Innov/Dev Ratio and then multiplied by 0.5 (by making an assumption that a developer spends about half their time on the project. This quantity is then multiplied by \$10,000 (assuming this to be the global average salary for a developer per month\footnote{\url{https://aijobs.net/salaries/developer-salary-in-2025/}}). By using projected lifetime growth, this approach also gives the lifetime builder cost of the project. It is also known as the ``supply-side'' value of the project. These costs vary significantly across the selected projects. The analysis is here is intended to be illustrative, showing how an understanding of the project life cycle is the first step in assessing lifetime costs and value generation. 

\subsection{Demand-side value}

Next, we evaluate the ``demand-side'' value of the project, i.e., what is the project worth to the users of the open-source software? To assess this value, we can use downloads as a proxy and if we can determine a value per download, then knowing the lifetime downloads enables an assessment of this value. However, demand-side valuation is complicated and subjective and these difficulties and a new approach to understanding the value of OSS is documented in \cite{chesbrough_measuring_2023}. 

\begin{figure*}
\centering
\caption{\label{fig:downloads_innov} \small The ratio of PyPi downloads to lines of code changed (additions and deletions). These ratios are determined by dividing the total downloads over the past 6 months by the total lines of code added and deleted over the same time frame. These ratios offer insight into the usefulness of projects to the community around them for the degree of effort invested by developers. 
}
\includegraphics[scale=0.7]{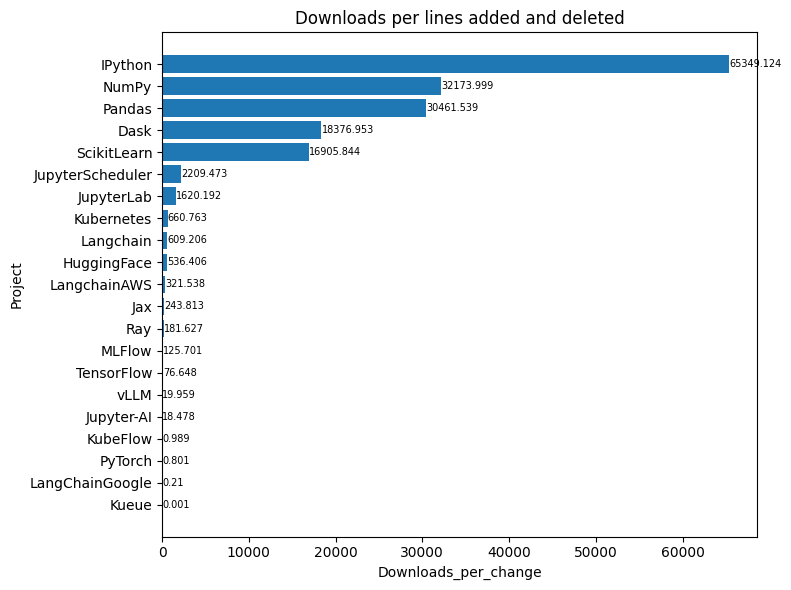}
{\footnotesize
\begin{tabular}{lllrrr}
\toprule
Project & Repo & PyPi package & 6mo downloads & 6mo changes & Ratio \\
\midrule
Dask &  dask/dask &  dask & 137036942 & 7457 & 18376.95 \\
HuggingFace &  huggingface/transformers &  transformers & 615182150 & 1146859 & 536.41 \\
IPython &  ipython/ipython &  ipython & 524949513 & 8033 & 65349.12 \\
Jax &  jax-ml/jax &  jax & 75853043 & 311112 & 243.81 \\
JupyterScheduler &  jupyter-server/jupyter-scheduler &  jupyter-scheduler & 121521 & 55 & 2209.47 \\
Jupyter-AI &  jupyterlab/jupyter-ai &  jupyter-ai & 897504 & 48572 & 18.48 \\
JupyterLab &  jupyterlab/jupyterlab &  jupyterlab & 254384699 & 157009 & 1620.19 \\
KubeFlow &  kubeflow/kubeflow &  kubeflow & 379218 & 383298 & 0.99 \\
Kubernetes &  kubernetes/kubernetes &  kubernetes & 519803832 & 786672 & 660.76 \\
Kueue &  kubernetes-sigs/kueue &  kueue & 753 & 677285 & 0.00 \\
LangchainAWS &  langchain-ai/langchain-aws &  langchain-aws & 38771056 & 120580 & 321.54 \\
Langchain &  langchain-ai/langchain &  langchain & 597444237 & 980693 & 609.21 \\
LangChainGoogle &  langchain-ai/langchain-google &  langchain-google & 22629 & 107567 & 0.21 \\
MLFlow &  mlflow/mlflow &  mlflow & 150869506 & 1200227 & 125.70 \\
NumPy &  numpy/numpy &  numpy & 3549725144 & 110329 & 32174.00 \\
Pandas &  pandas-dev/pandas &  pandas & 2772426479 & 91014 & 30461.54 \\
PyTorch &  pytorch/pytorch &  pytorch & 1179989 & 1473048 & 0.80 \\
Ray &  ray-project/ray &  ray & 203893216 & 1122592 & 181.63 \\
ScikitLearn &  scikit-learn/scikit-learn &  scikit-learn & 898376562 & 53140 & 16905.84 \\
TensorFlow &  tensorflow/tensorflow &  tensorflow & 148662915 & 1939560 & 76.65 \\
vLLM &  vllm-project/vllm &  vllm & 23646706 & 1184765 & 19.96 \\
\bottomrule
\end{tabular}
}
\end{figure*}

The growth model (lines added and deleted) offers a lifetime forecast, as shown in Table \ref{tab:valuation}. If we have an estimate of the number of downloads per lines edited for each project from existing data, we can estimate lifetime downloads. The ratio of PyPi downloads to lines of code changed varies across projects, with the popular projects having very high values compared to the less popular ones as shown in Figure \ref{fig:downloads_innov}. 

This information may be used to estimate the demand-side value. For example, the ratio for Pandas is 30462 (Figure \ref{fig:downloads_innov}). Since this project has a lifetime growth in code of 5.24 million lines edited (Table \ref{tab:valuation}), the lifetime downloads are expected to be 160 billion. Put a value on each download to determine what the project is worth. 

Another simpler calculation may be as follows. We know that Pandas has 2,772,426,479 downloads in six months (Figure \ref{fig:downloads_innov}), and we can extrapolate this over the remaining life of the project, which is 153 months (\ref{tab:valuation})), i.e., $2772426479 \times 153/6 = 70,696,875,215$ downloads to go. The devil is in the assumptions, and whatever these may be, they should be consistently applied across time and projects. 

To be clear, these are only suggested approaches to lifetime project value based on first estimating lifetime developer engagement and projected endogenous growth. The number of lines of code committed by a developer per month can vary widely depending on various factors, but typical ranges may vary depending on which types of coders are being considered. Anecdotal estimates suggest that professional developers tend to commit between 10-50 lines of code per day.\footnote{\url{https://softwareengineering.stackexchange.com/questions/40100/how-many-lines-of-code-can-a-c-developer-produce-per-month}} \footnote{ \url{https://stackoverflow.com/questions/966800/mythical-man-month-10-lines-per-developer-day-how-close-on-large-projects}} This translates to roughly 200--1000 lines of code per month, assuming 20 working days.\footnote{\url{https://news.ycombinator.com/item?id=37040552}} Then of course, several factors influence the number of lines of code committed, such as project complexity, development phase, individual coding style, team size and collaboration, and code review processes. Admittedly, lines of code amended is a fuzzy measure of productivity because some days may involve writing hundreds of lines of code, while others might focus on debugging or refactoring with minimal new code. Experienced developers often write fewer, more efficient lines of code. Impact may also be measured in other ways such as code quality and the impact on project goals, which may vary a lot across projects, and are quite subjective. 

\section{Concluding Discussion}
\label{sec:conclusions}

The paper adapts the endogenous growth model of \cite{romer_endogenous_1990} to the dynamic evolution of code growth levels and developer involvement in open-source software projects. Developer engagement is modeled using the \cite{bass_new_1969} model, which fits the data well, suggesting that a product life cycle model is adaptable to modeling open-source project life cycles. This paper's  system of dual ODEs for growth and developer engagement is solved and calibrated to data from GitHub repositories. Phase diagrams enable an assessment of the maturity of the project, its future trajectory, and the elasticities of growth and contributors on these dynamics. 

We are able to speculate when developer engagement will drop below a single developer-month, without any further exogenous changes, signaling a point of low usage and activity.  We applied this to some popular projects with interesting results. To the extent that these life cycle projections are within acceptable levels of accuracy, they may be used to forecast how long we may expect growth in and depend upon  an open-source project (see Figures \ref{fig:all_lifecycles}, \ref{fig:model_fits1}, and \ref{fig:model_fits2}). This complements the idea of underproduction risk noted in \cite{champion_underproduction_2021}, \cite{gaughan_engineering_2024}.

Experimentation across a few different systems of ODEs suggest some regularities, namely that growth feeds on itself, like a flywheel, and once this is in motion, additional developers do not matter as much and efficiency and downloads per unit of developer effort improves. This explains why open-source projects are maintained by a small group of core developers, noted in \cite{droesch_measuring_2020}. As projects mature and taper off in developer engagement as shown in Figure \ref{fig:downloads_innov}, we may detect successful mature projects as those that have a high download to developer effort ratio.

Finally, we are able to use the dynamical equations for developer engagement and cumulative growth to support illustrative valuation models for open-source projects. Supply-side valuation for 10 chosen projects ranged from the tens of millions to the hundreds of millions. Demand-side valuation can be imputed from lifetime growth using the ratio of downloads to lines of code changed. These are initial approaches to understanding the life cycle of open-source software projects. Future work will expand the number of projects analyzed. The analysis of strategic investments by tech companies in these projects and its benefits may be possible with minor additions to this model.

\bibliographystyle{chicago} 
\bibliography{OpenSource}

\appendix

{\it Solving the initial differential equations for project growth and developer engagement.} See Section \ref{sec:endogenous_growth_model}.

The dynamics of growth level $A$ are described by equation (\ref{dotA}), reproduced here.    
\begin{equation}
{\dot A} = \frac{dA}{dt} = \gamma L^{\lambda} A^{\phi} \end{equation}
The equation for developer engagement (labor) is reproduced here as well:
\begin{equation}
{\dot L} = \frac{dL}{dt} = n L  \label{dotL}
\end{equation}
With initial conditions:
$$
L(0) = L_0, \quad A(0) = A_0
$$

The solution to this system of equations is as follows:
\begin{enumerate}
\item First, let's solve the equation for Labor ($L$) since it's independent of $A$:
$$ \dot{L} = nL $$
This is a simple separable differential equation:
$$ \frac{dL}{dt} = nL $$
$$ \frac{dL}{L} = n \; dt $$
$$ \int \frac{dL}{L} = \int n \; dt $$
$$ \ln|L| = n \cdot t + C $$
Using the initial condition $L(0) = L_0$:
$$ L(t) = L_0e^{nt} $$  
\item 
Now we can substitute this solution into the equation for A:
$$ \dot{A} = \gamma(L_0e^{nt})^\lambda A^\phi $$
This is a Bernoulli differential equation in A:
$$ \frac{dA}{dt} = \gamma L_0^\lambda e^{\lambda nt}A^\phi $$
\item To solve this, let's use the substitution $u = A^{1-\phi}$:
$$ \frac{du}{dt} = (1-\phi)A^{-\phi}\frac{dA}{dt} $$
Substituting:
$$ \frac{du}{dt} = (1-\phi)\gamma L_0^\lambda e^{\lambda nt} $$
\item 
Integrating both sides:
$$ u = \frac{(1-\phi)\gamma L_0^\lambda}{\lambda n}e^{\lambda nt} + C $$
\item 
Substituting back $u = A^{1-\phi}$:
$$ A^{1-\phi} = \frac{(1-\phi)\gamma L_0^\lambda}{\lambda n}e^{\lambda nt} + C $$
\item 
Using the initial condition $A(0) = A_0$:
$$ A_0^{1-\phi} = \frac{(1-\phi)\gamma L_0^\lambda}{\lambda n} + C $$
$$ C = A_0^{1-\phi} - \frac{(1-\phi)\gamma L_0^\lambda}{\lambda n} $$
\item 
Therefore, the complete solution for both processes is:
{\small
\begin{eqnarray}
L(t) &=& L_0e^{nt} \label{Lsol}\\
A(t) &=& \left(\frac{(1-\phi)\gamma L_0^\lambda}{\lambda n}e^{\lambda nt} + A_0^{1-\phi} - \frac{(1-\phi)\gamma L_0^\lambda}{\lambda n}\right)^{\frac{1}{1-\phi}}  \label{Asol}
\end{eqnarray}
}
This solution is valid for $\phi \neq 1$, $n>0$. If $\phi = 1$, the equation would need to be solved differently. Also, for $n=0$, the solution is shown in equation (\ref{n0}). To check the solution by differentiation, set equation (\ref{Asol}) to $A(t)=u(t)^{1/(1-\phi)}$. so that $\frac{dA}{dt}=\frac{1}{1-\phi} \cdot u^{\frac{\phi}{1-\phi}} \cdot \frac{du}{dt}$, and then proceed to recover equation (\ref{dotA}) in a few steps of algebra. 
\end{enumerate}
The overall solution shows that: (a) Labor grows exponentially at rate $n$. (b) The growth level A grows as a function of both the initial conditions and the parameters $\gamma$, $\lambda$, $\phi$, and $n$.

\end{document}